\documentclass[aps,twocolumn,superscriptaddress,showpacs,showkeys,pra,amssymb, amsmath]{revtex4-1}
\usepackage{amsmath,latexsym,amssymb,amscd}
\usepackage{graphicx}
\usepackage{amsmath}
\usepackage{latexsym} 
\usepackage{amssymb} 

\newtheorem{theorem}{Theorem}
\newtheorem{thm}[theorem]{Theorem}

\newtheorem{lem}[theorem]{Lemma}

\begin{document}
\title{Statistical Tests and Confidential Intervals as Thresholds for Quantum Neural Networks}
\date{\bf \today}

\author{Do Ngoc Diep}
\address{TIMAS, Thang Long University, Nghiem Xuan Yem Road, Hoang Mai district, Hanoi, Vietnam}
\address{Institute of Mathematics, Vietnam National Academy of Science and Technology, 18 Hoang Quoc Viet road, Cau Giay district, 10307 Hanoi, Vietnam}

\email{diepdn@thanglong.edu.vn}

\date{\textbf{\today}}

\begin{abstract}  
\textbf{Abstract}

Some basic quantum neural networks were analyzed and constructed in the recent work of the author \cite{dndiep3}. In particular the  Least Quare Problem (LSP) and the Linear Regression Problem (LRP) was discussed. 
In this second paper we continue to analyze and construct the least square quantum neural network (LS-QNN), the polynomial interpolation quantum neural network (PI-QNN), the  polynomial regression quantum neural network (PR-QNN) and chi-squared quantum neural network ($\chi^2$-QNN). We use the corresponding solution or tests as the threshold for the corresponding training rules.
\end{abstract}
\maketitle
\section{Introduction}

The classical machine learning (ML) \cite{wiebeetal},\cite{hinton} theory was created in 1950, but only  9 years later  in 1959 Arthur Samuel gave a definition being ``.... computers learning without being explicitly programmed''.
 It should understand that the functions (inputs-outputs) are deduced from a set of training data. The classical ML is characterized with: 1) \textit{supervised learning}, i.e. classes of inputs corresponds to different classes, (2) \textit{unsupervised learning}, i.e. the large data are summarized into a few stereotypes, and (3) \textit{reinforcement learning}, i.e. rewards, reinforce the current strategy. Normally the classical MLs are working with \textit{big data}, see \cite{schuldetal},\cite{ezhovventura},.

The quantum Machine Learning (QML) are characterized by using quantum computing into the theory. One uses the ordinary interpretation of qubits, 1-qubit quantum gates, such as the Pauli matrices, etc. 
$$\mathbf 1 = -\fbox{Id}- \sim \begin{pmatrix} 1 & 0\\ 0 & 1 \end{pmatrix},$$  $$X= -\fbox{X}-\sim \begin{pmatrix} 0 & 1 \\ 1 & 0\end{pmatrix},$$ $$Y = -\fbox{Y}-\sim \begin{pmatrix} 0 & -i \\ i & 0 \end{pmatrix},$$ $$Z = -\fbox{Z}-\sim\begin{pmatrix} 1 & 0\\ 0 & -1 \end{pmatrix},$$ then the 2-qubit gates like $$\mathrm{XOR} = -\fbox{XOR}- \sim \begin{pmatrix} 1 & 0 &  0 & 0\\
0 & 1 & 0 & 0\\ 0 & 0 & 0 &1\\ 0 & 0 & 1 & 0\end{pmatrix},$$  $$\mbox{SWAP} = -\fbox{SWAP}- \sim  \begin{pmatrix} 1 & 0 & 0 & 0\\ 0 & 0 & 1 &0\\ 0 & 1 & 0 & 0\\ 0 & 0 & 0 & 1 \end{pmatrix},$$ and finally,  Measurements $$\mathrm{M}=-\fbox{M}-. $$

One uses the quantum algorithms to solve the ML problems with quantum computing. The most important ingredients in QML are:  - choices of  \textit{training sets}, i.e. finite sets of given vectors in order to then find some value corresponding to another input, - \textit{pattern completion}, i.e. adding missing informations to incomplete inputs, and  - \textit{associative memory}, i.e. retrieving  stored memory vectors upon an input.

This paper is the second part of the paper \cite{dndiep3}, in which we continue to treat the cases of polynomial regression with the high probability region bounds used as the corresponding thresholds.
In Section II, we analyze the comceptions of classical aritificial neural networks (ANN) and quantum neural networks (QNN). The next section III is devoted to the problem of training the least square quantum neural networks (LS-QNN), like the least square interpolation, the general polynomial regression quantum neural network (PR-QNN) and the chi-squared test training ($\chi^2$-QNN) in the next section IV.
We look at the problem of least square problem (LSP) solution of the general polynomial regression and propose to use the quantum Gauss-Jordan Elemination (GJE) Code to solve the LSP equation. This let us to make the network works outperform the classical approaches. 
The paper is finished with a conclusion in section IV and the last section V is our acknowledgments.

\section{Quantum Neural Networks}
Following the model of Deutsch, a quantum neural network $\mathrm{QNN}(s,d)$ is a set of all quantum circuit of seize $s$ and depth $d$ with thresholds bounded by $w$. Quantum gates are interconnected by wires, preserve the sources and sink gates (measured the qubits and removed the entanglements with the maining qubits. Examples of QNNs are the implementation of NAND gate, dissipative $D(m,\delta)$ and sink gates.

A threshold circuit is a boolean function $Th^{n,\Delta}: \mathbb Z_2^n \to \mathbb Z_2$ of $n$ integral variables $x_1,\dots,x_n$ such that $Th^{n,\Delta}(x_1,\dots,x_n) = 1$ if and only if $\sum x_i \geq \Delta$. The class $TC(s(n),d(n))$ of threshold circuits of size $s(n)$ and depth $d(n)$, weighted by weight bound $w$ can be approximated by elementary functions.  

An equality threshold circuit is a boolean function $Et^n_{w_1,\dots,w_n}: \mathbb Z_2^n \to \mathbb Z_2$ of $n$ integral variables $x_1,\dots,x_n$ such that $Et^n_{w_1,\dots,w_n}(x_1,\dots,x_n) = 0$ if and only if $\sum x_i =0$. The class $EC(s(n),d(n))$
 of equality threshold circuits of size $s(n)$ and depth $d(n)$, weighted by weight bound $w$ can be approximated by elementary functions.  
 
 It was proven that $TC(s(n),d(n)) \subseteq EC(O(s^2(n),2d(n))$ of weight bound $O(s(n))$ and $TC(s(n),d(n)) \subseteq EC(O(s^2(n),d(n)+1)$ of weight bound $O(s^2(n))$. And finally, $EC(s(n),d(n)) \subseteq QNN(O)d(n).\log s(n)), 2d(n))$ of precision $O(\log w + d(n)\log s(n))$.
 (Theorem 4.6 from \cite{guptazia}).

The question is whether  a QNN can be implemented on Quantum Turing Machine (QTM) (Church-Turing Thesis) is difficult to answer: Quantum computing showed that it is No, but physicists speculate that it is Yes.

\section{Least Square QNN and Polynomial Regression QNN}
First we remind that many problem, including the least squared problem and polynomial interpolation problems are reduced to solving systems of linear equations. In the previous work \cite{dndiep1} we had showed that the Gauss-Jordan elimination procedure is consisting of an application of searching the pivot columns, which is reduced to use the Gover's Search Algorithm and by the way necessary arithmetic operations over rows. The following lemma \cite{dndiep3} is fundamental in many problems of namely the least square or the polynomial interpolation quantum neural networks.
\begin{lem} \label{Lemma 1} The quantum Gauss-Jordan Elimination Code  can be implemented in QNN.
\end{lem}
Let us consider the polynomial $f(\mathbf x) = \sum_{|\alpha|=0}^{N} a_\alpha \mathbf x^\alpha$ of degree $N$ on $n$ variables, with unkown coefficients $a_\alpha$, those we want to inpterpolate, and let
$(\{\mathbf x_{(j)}^\alpha\}\}_{|\alpha|=0}^{N},y_j)$, $\alpha =(\alpha_1,\dots,\alpha_n)$, $|\alpha|= \alpha_1+\dots+\alpha_n\leq N$ be the $N+1$ interpolating points of the polynomial, $x= (x_1,\dots x_{n})$ be the unkown variables, $\mathbf x_{(j)}^\alpha = \Pi_{i=1}^n {x_{i,(j)} }^{\alpha_i}, j=0,\dots, N$. 
The system of interpolating equations is a system of $N+1$ equation on $N+1$ \textsl{unknown variables} $a_\alpha, |\alpha|= 0,\dots,N$: 
$$ f(\mathbf x_{(j)}) = \sum_{|\alpha|=0}^{N} a_\alpha \mathbf x_{(j)}^\alpha= y_{j}; j=0,\dots, N.$$ 
The determinant of the system is of the Vandermonde type and of size $(N+1)\times(N+1)$ 
$$|A| = \left|   
\begin{matrix} 1 & x_{(0)}^{(1,\dots,0)} & \dots & x_{(0)}^{(0,\dots,N)} \\ 
1 & x_{(1)}^{(1,\dots,0)} & \dots & x_{(1)}^{(0,\dots,N)} \\ 
\dotfill &\dotfill & \dotfill &\dotfill \\
1 & x_{N}^{(1,\dots,0)} & \dots & x_{N}^{(0,\dots,N)} 
\end{matrix}
\right| ,$$
then the system can be written as 
$$A^\dagger A\mathbf [a_\alpha] = A^\dagger [\mathbf y_j]. \eqno{(3.1)} \label{1}$$ 

The matrix of the system is nonvanishing if the interpolating points are in a generic position. 
In that case the solution of the system is 
$[a_\alpha]_{|\alpha=0}^{N} = (A^\dagger A)^{-1} A^\dagger \mathbf b$, where $\mathbf b=[\mathbf y_j]_{j=0}^N$.

In general case the matrix can not be invertible, but the system is consistent. Based on Lemma \ref{Lemma 1}, we can use the Gauss-Jordan elimination procedure on quantum neural networks to find out a basis of the null-space of the augmented matrix of the system (3.1).  Let $(A^\dagger A)_{psi}^{-1}$ be the Moore-Penrose pseudoinverse of $A^\dagger A$,
then the solution to the interpolation problem is
$[a_\alpha] = (A^\dagger A)^{-1}_{psi} A^\dagger b$.,
where $\mathbf b = \mathrm{proj}_{col (A^\dagger A)} A^\dagger [\mathbf y_j]$. 

The general interpolated solution is 
$$ \hat f(\mathbf x_{(j)}) = \sum_{|\alpha|=0}^{N} a_\alpha \mathbf x_{(j)}^\alpha= \hat y_{j}; j=0,\dots, N. \eqno{(3.2)}$$ 

We have therefore the following result
\begin{thm}
The Least Square Quantum Neural Network (LS-QNN) and Polynomial Interpolation Quantum Neural Networks(PI-QNN) are  implementable on QNN, with complexity $O(\sqrt{N})$.
\end{thm}

We now apply the Least Square Method to the problem of (general) regression (GRP).
Let us remind that the Grover's Searh Code can be implemented in QNN because the basic step is to repeatedly use the XOR quantum network gate \cite{dndiep3}.
The method of QGJE \cite{dndiep1} is based on use of the Quantum Grover's Search to find the pivot columns in the matirx $A^\dagger A$. 

We have therefore the following result
\begin{thm}
The Polynomial Regression Quantum Neural Network (PR-QNN) is implementable, i.e. the GRP can be solved by a QNN, with complexity $O(\sqrt{N})$.
\end{thm}
Let us analyze how to train the GRP code in QNN. With the above interpolating quantum code, we can divide the data $y_j$ into to treatment: regression treatment $Y_{regr}= [\hat{y}_j]$ and residual treatment $Y_{resid}= [y_j - \hat y_j]$, where $$\hat y_j = f_{regr}(\mathbf x_{(i)})=\hat f(\mathbf x_{(j)}).\eqno{(3.3)}$$ Let us denote by $$F = \frac{MS_{regr}}{MS_{resid}} = \frac{\frac{(r^2SS_Y}{1}}{\frac{(1-r^2)SS_{resid}}{N-2}}= \frac{(N-2)r^2}{1-r^2},\eqno{(3.4)}$$ where $r$ is the Pearson correlation, $r=Cor(X,Y)$. We may fix a level $\alpha$ of explained proportion of variance and define the $F$-ratio $F_{(1,N-2),\alpha}$. Therefore we define the \textit{training threshold} as if the $F$-ratio is in the high probability $1-\alpha$ region $$F< F_{(1,N-2),\alpha}.\eqno{(3.5)}$$ \hfill$\Box$

\section{Chi-Squared QNN}
In the nonparametric statistics, the $\chi^2$-test plays important roles in many problems like contingency tables, homogeneity, .......
Let use conside the corresponding quantum code in QNN.  Denote by $\mathbf e = [e_{ij}]_{n\times r}$ be a contingency matrix of expected values $e_{ij}$.
The random distribution $X= [x_{ij}]$ is a matrix of size $n \times r$. The degree of freedom is $$df_X = \begin{cases} (n-1) \times (r-1), &\mbox{ if } r>1\\  (n-1) & \mbox{ if } r=1\end{cases}$$. The chi-squared statistic is of form
$$\chi^2_X = \sum_{i=1}^n\sum_{j=1}^r \frac{(x_{ij} - e_{ij})^2}{e_{ij}}. \eqno{(4.1)}$$
Our aim is to implement the $\chi^2$-test in a QNN and use the chi-sqaured test as the rule of training.

\begin{thm}
The Chi-Squared Quantum Neural Network ($\chi^2$-QNN) is implementable, , i.e. the $\chi^2$-test can be solved by a QNN, with complexity $O(\sqrt{n-1) \times (r-1)})$.
\end{thm}
Indeed, the high probability $1-\alpha$ region is $$\chi_X^2 < \chi_{df,\alpha}\eqno{(4.2)}$$ for a fixed $\alpha$-level of confidence and\\ the \textit{training rule} is to sink the network if the constraint is faile to be satisfied. If the constraint holds, it passes to the next layer of QNN.  \hfill$\Box$

\section{Conclusion}
We implemented the quantum neural networks: the least square quantum Neural Network (LS-QNN) and the polynomial interpolation quantum neural networks (PI-QNN), the Polynomial Regression Quantum Network (PR-QNN) and the Chi-Squared Quantum Neural Network ($\chi^2$-QNN). The training rules are provided with the corresponding test from Statistics.
  
\section{Acknwledgments}
The authors express their sincerle thanks to Professor K. Nagata, Professor G. Resconi, Professor T. Nakamura, Professor S. Heidari and the referees for careful reading the manuscripts and valuable comments.


\begin{thebibliography}{xxx}


\bibitem{wiebeetal}{\sc N. Wiebe, A. Kapoor, K. Swore}, {Quantum deep learning}, arXiv:1412.3489v2[quant-ph], 2015.

\bibitem{hinton}{\sc G. Hinton}, {\it A Practical Guide to Training Restricted Boltzmann Machines}, 
Department of Computer Science, University of Toronto. 



\bibitem{schuldetal}{\sc M. Schuld, I. Sinayskiy, Petruccione}, {\it An introduction to quantum machine learning}, arXiv:1409.3097v1[quant-ph]2014.


\bibitem{ezhovventura}{\sc A. A. Ezhov,  D. Ventura}, {\it Quantum neural networks}, in Future \textit{Directions for Intelligent Systems and Information Science},  N. Kasabov (ed.), Physica-Verlag, pp. 213-235, 2000.




\bibitem{schuldsinayskiypetruccione}{\sc M. Schuld, I. Sinayskiy, F. Petruccione}, {\it Prediction by linear regression on a quantum computer}, arXiv:1601.07823v2[quant-ph], 2016.

\bibitem{dndiep1}{\sc D. N. Diep, D. H. Giang, N. V. Minh},  {\it Quantum
Gauss-Jordan elimination and simulation ofnaccounting principles
 on quantum computers}, Inter. J. of Theor. Physics, 56(2017), No 6, 1948-1960.

\bibitem{dndiep2}{\sc K. Nagata, S. K. Patro, H. Geurdes, S. Heidari, D. N. Diep, T. Nakamura}, {\it Various New Forms of the Bernstein-Vazirani Algorithm Beyond Qubit Systems},  Asian J. Math. \& Phys., 3, No 1(2019) 1-12.

\bibitem{dndiep3}{\sc D. N. Diep}, {\it Some quantum neural networks}, Intl. J. Theor. Phys. (to appear).


\bibitem{guptazia}{\sc S. Gupta, R.K. P. Zia}, {\it  Quantum Neural Network}, Journal of computer and system sciences, 63(2001), 355-383.

\end{thebibliography}
\end{document}